\documentclass[sigconf]{acmart}

\renewcommand\footnotetextcopyrightpermission[1]{} 
\setcopyright{none}           
\makeatletter
\def\@copyrightspace{}        
\makeatother

\acmDOI{}                     
\acmISBN{}                    
\acmPrice{}                   
\acmConference[]{}{}{}        

\usepackage{fancyhdr}
\fancypagestyle{firstpage}{
  \fancyhf{}
  
  \fancyfoot[L]{\footnotesize 
    \begin{minipage}[t]{0.45\textwidth} 
    Accepted at the International Conference on Evaluation and Assessment in Software Engineering (EASE 2026).
    \end{minipage}
  }
  \fancyfoot[C]{\footnotesize \thepage} 
}

\AtBeginDocument{%
  \providecommand\BibTeX{{%
    Bib\TeX}}}

\acmConference[EASE 2026]{Accepted to The 30th International Conference on Evaluation and Assessment in Software Engineering}{9–12 June, 2026}{Glasgow, Scotland, United Kingdom}

\usepackage[most]{tcolorbox}
\usepackage{amsmath,amssymb,amsfonts}
\usepackage{algorithmic}
\usepackage{graphicx}
\usepackage{textcomp}
\usepackage{xcolor}
\usepackage{url}
\usepackage{comment}
\usepackage[normalem]{ulem}
\def\BibTeX{{\rm B\kern-.05em{\sc i\kern-.025em b}\kern-.08em
    T\kern-.1667em\lower.7ex\hbox{E}\kern-.125emX}}

\begin{document}
\title{Mining the YARA Ecosystem: From Ad-Hoc Sharing to Data-Driven Threat Intelligence}

\author{Esteban Dectot--Le Monnier de Gouville}
\email{esteban.dectot@polymtl.ca}
\orcid{0009-0004-6262-0552}
\affiliation{%
  \institution{Polytechnique Montréal}
  \city{Montréal}
  \state{Québec}
  \country{Canada}
}

\author{Mohammad Hamdaqa}
\email{mhamdaqa@polymtl.ca}

\affiliation{%
  \institution{Polytechnique Montréal}
  \city{Montréal}
  \country{Québec}
  \country{Canada}}

\author{Moataz Chouchen}
\email{moataz.chouchen@concordia.ca}
\affiliation{%
  \institution{Concordia University}
  \city{Montréal}
\state{Québec}
  \country{Canada}
}
\renewcommand{\shortauthors}{Dectot--Le Monnier de Gouville, et al.}

\begin{abstract}
YARA has established itself as the de facto standard for "Detection as Code," enabling analysts and DevSecOps practitioners to define signatures for malware identification across the software supply chain. Despite its pervasive use, the open-source YARA ecosystem remains characterized by ad-hoc sharing and opaque quality. Practitioners currently rely on public repositories without empirical evidence regarding the ecosystem’s structural characteristics, maintenance and diffusion dynamics, or operational reliability. We conducted a large-scale mixed-method study of 8.4 million rules mined from 1,853 GitHub repositories. Our pipeline integrates repository mining to map supply chain dynamics, static analysis to assess syntactic quality, and dynamic benchmarking against 4,026 malware and 2,000 goodware samples to measure operational effectiveness. We reveal a highly centralized structure where 10 authors drive 80\% of rule adoption. The ecosystem functions as a "static supply chain": repositories show a median inactivity of 782 days and a median technical lag of 4.2 years. While static quality scores appear high (mean = 99.4/100), operational benchmarking uncovers significant noise (false positives) and low recall. Furthermore, coverage is heavily biased toward legacy threats (Ransomware), leaving modern initial access vectors (Loaders, Stealers) severely underrepresented. These findings expose a systemic "double penalty": defenders incur high performance overhead for decayed intelligence. We argue that public repositories function as raw data dumps rather than curated feeds, necessitating a paradigm shift from ad-hoc collection to rigorous rule engineering. We release our dataset and pipeline to support future data-driven curation tools.
\end{abstract}


\received{26 January 2026}
\received[accepted]{13 March 2026}

\maketitle
\thispagestyle{firstpage}

\section{Introduction}

YARA is a rule-based pattern-matching language widely used in malware detection and incident response. Its flexibility, portability, and efficiency have established it as the \textit{de facto} standard for encoding detection logic across both open-source and commercial environments~\cite{virustotalYARA}. Originally designed for digital forensics, YARA has expanded with the rise of DevSecOps into CI/CD pipelines, embedding malware detection directly into modern software delivery workflows~\cite{Malik2025ShiftLeft, YaraHunter}. 

\begin{figure}[ht!]
    \centering
    \includegraphics[width=.9\linewidth]{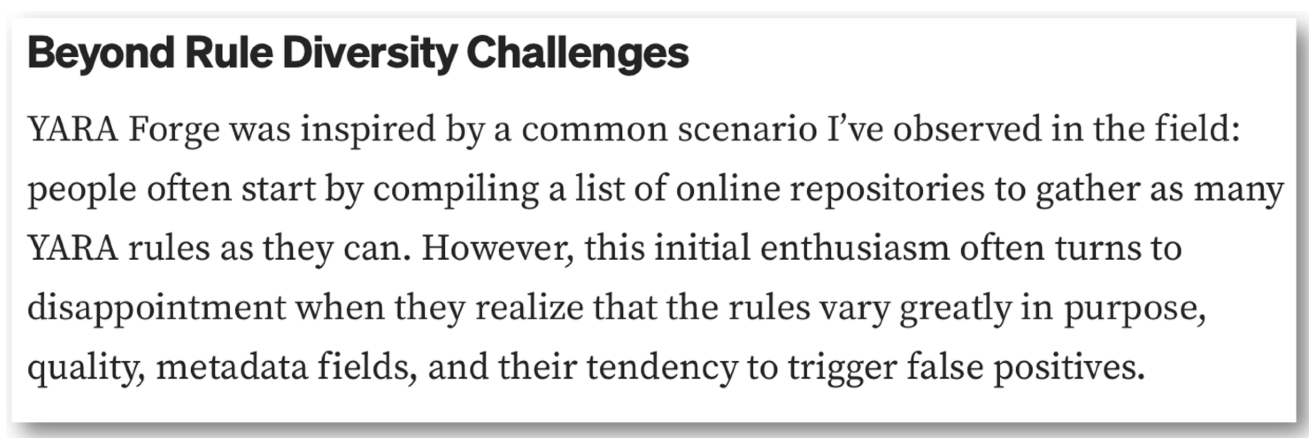}
    \caption{Quote from Florian Roth introducing YARA Forge on Medium~\cite{roth2023yaraforge}}
    \label{fig:quote}
\end{figure}

Today, the language is supported by a massive ecosystem: it is integrated into major security platforms (e.g., VirusTotal, Cisco, Kaspersky) and fueled by a sprawling community of open-source threat intelligence sharing. Despite this ubiquity, the open-source YARA ecosystem itself remains poorly characterized. As noted by Florian Roth, creator of YARA Forge, practitioners often rely on ad-hoc rule sharing and maintenance, leading to frustration and inconsistent detection results (see Figure~\ref{fig:quote}). While early attempts at characterization exist, they address only narrow aspects—such as automated rule generation or syntax checking~\cite{roth2023yaraforge, Patil2025YARA}—yet no comprehensive, data-driven characterization of the ecosystem's structure, evolution, and operational reliability has been conducted.

This paper presents the first large-scale, systematic characterization of the open-source YARA ecosystem. \textbf{We adopt a mixed-method engineering research design}, combining large-scale repository mining with operational benchmarking. Repository mining method permits to capture the socio-technical dynamics of rule sharing at scale. As established in foundational OSS research by Mockus et al. \cite{mockusOSS}, the distributed nature of open-source collaboration means that 'virtually all information on the OSS project is recorded in electronic form'. Consequently, analyzing these archival traces is necessary to reconstruct the history, structure, and quality of the ecosystem without the sampling bias inherent to surveys or interviews. Repository mining allows us to map the ecosystem's dynamics (creation, reuse, maintenance), but it cannot assess the utility of the shared rules. Therefore, we integrate a dynamic benchmarking phase to contrast the theoretical quality (syntax) against operational reality (detection). To this end, we analyze 8.4 million rules from 1,853 GitHub repositories to answer three fundamental research questions:

\begin{itemize}
    \item \textbf{RQ1 (Ecosystem Dynamics): How can we characterize the open-source YARA ecosystem?} \\
    We study the ecosystem’s dynamics by investigating three complementary dimensions:
    \begin{itemize}
        \item \textit{RQ1.1 (Structure):} \textbf{How are open-source YARA rules distributed and replicated across GitHub repositories?} \\
        We analyze the distribution of unique rule logic using fuzzy hashing to quantify redundancy and assess whether rule sharing is broadly distributed or concentrated in a small number of hubs.

        \item \textit{RQ1.2 (Author Influence):} \textbf{Who contributes YARA rules, and how concentrated is their ecosystem impact?} \\
        We examine contributor influence by comparing productivity to downstream adoption, quantifying the degree of concentration in rule creation and reuse across repositories.

        \item \textit{RQ1.3 (Evolution):} \textbf{How do YARA rules evolve over time across repositories (maintenance and update propagation)?} \\
        We characterize rule evolution by measuring \textbf{Technical Lag}---the delay between a rule’s first appearance and its adoption in other repositories---to assess whether shared repositories behave as continuously updated feeds or static archives.
    \end{itemize}
    \item \textbf{RQ2 (Operational Reliability): What is the operational reliability of public YARA rules?}
    We assess the gap between theoretical quality and practical utility. We combine static analysis (using \textit{yaraQA}) with dynamic benchmarking against 1,910 benign and malicious samples. We find that high static quality scores are commonplace but unrelated to operational effectiveness; syntactically perfect rules frequently exhibit high false-positive rates or fail to trigger on relevant threats, highlighting the limitations of current static validation tools.
    \item \textbf{RQ3 (Threat Relevance): To what extent do public YARA rules cover different malware families?}
    We examine whether the ecosystem addresses the modern threat landscape or remains biased toward legacy threats. Using a semi-automated taxonomy classifier, we map rules to specific malware families and study the first appreance of a rule covering it. Our analysis uncovers systematic blind spots: while high-profile threats like Ransomware are over-represented, critical initial access vectors—specifically Loaders and Stealers—are severely under-detected, leaving a gap in the early stages of the kill chain.
\end{itemize}


\textbf{Contribution.} This paper moves beyond anecdotal evidence by presenting a data-driven map of the open-source YARA landscape. Our results provide empirical evidence that the rule-sharing supply chain is highly centralized and slow to propagate updates, indicating that many public repositories function more as static rule archives than curated intelligence feeds.

\textbf{Open science.} To facilitate the replication and the extension of our study, we share our data and scripts in our replication package \cite{YARA_Ecosystem_Replication_2026}.

\section{Background}
YARA (Yet Another Recursive Acronym) is a versatile pattern-matching tool that has evolved beyond its security roots to support broader software engineering applications \cite{virustotalYARA}. Unlike traditional signature systems, YARA enables expressive rules that combine metadata, string patterns, and boolean logic to identify specific characteristics in digital artifacts. As shown in Figure~\ref{fig:yara_example}, each rule includes metadata for documentation, string definitions for pattern matching (including regular expressions), and conditional logic specifying match criteria.
\begin{figure}[htbp]
\begin{verbatim}
rule Time_Wasting_Virus_TW1_2025 {
    meta:
        description = "Detects the Time-Wasting Virus"
        author = "Anonymous Author"
        threat_level = "Critical (for productivity)"
    strings:
        $later = "I'll fix it later"
        $cat = "youtu.be/shorts/"
        $coffee = "Just one more coffee"
    condition:
        any of them
}
\end{verbatim}
\caption{Fictive example of a YARA rule}
\label{fig:yara_example}
\end{figure}
Originally developed for malware analysis, YARA has gained adoption across the software industry due to its flexibility in defining custom detection criteria. The example above illustrates how YARA can identify productivity anti-patterns in codebases, demonstrating its applicability beyond traditional security use cases. This particular rule detects common indicators of technical debt and procrastination in development environments.

YARA’s linear-time complexity and optimized pattern matching make it well-suited for integration into software development workflows. It has been widely adopted by major technology companies and integrated into various ecosystems through language bindings and CI/CD plugins. Its capability to process both textual and binary files supports applications ranging from code quality analysis to build artifact verification.
The open-source nature of YARA has fostered a rich ecosystem of community-maintained rule repositories, generation tools, and pipeline integrations \cite{awesomeYara}. This ecosystem enables software teams to implement customized analysis solutions that detect code patterns, enforce standards, and identify issues early in development. YARA’s flexibility in defining “matches” allows organizations to codify domain knowledge and best practices into automated, reusable checks across projects.

\section{Related Works}
Current research on signature-based detection can be categorized into three streams: the microscopic analysis of individual rules, the automated generation of signatures, and the macroscopic analysis of software ecosystems. While the first two are well-represented in YARA literature, our work introduces methodologies from the third stream—specifically empirical ecosystem analysis and DevSecOps—to the domain of threat detection.

Foundational studies on YARA rule quality have primarily focused on the intrinsic properties of effective signatures, establishing the critical dimensions of \textit{robustness} (detecting variants) and \textit{looseness} (false positive propensity)~\cite{canfora2020robustness}. This framework has been codified by static analysis tools like yaraQA~\cite{neo23x0_yaraqa}, which enforce syntactic best practices. Parallel research has addressed the computational cost of scanning, demonstrating how inefficient regular expressions degrade performance~\cite{regeciova2023yara} and proposing sub-linear search architectures like YARIX~\cite{Brengel2021YARIX} to handle massive corpora.
Simultaneously, to overcome the manual bottleneck of signature creation, recent approaches have leveraged Large Language Models (LLMs). Tools like RuleLLM~\cite{Zhang2025Automatically} and other generative methods~\cite{raff_automatic_2020} report high precision in controlled experiments.
\textit{\uline{However, these streams implicitly view the ruleset as a static, trusted oracle. Our study challenges this by shifting focus from the syntactic correctness of a single rule to the ecosystem-level dynamics and operational reliability of shared rules.}}

While characterizing ecosystem dynamics in rule repositories is novel in malware research, it is a well-established discipline in empirical software engineering. Extensive studies on open-source ecosystems (e.g., npm, PyPI, Maven) have defined metrics to quantify ecosystem sustainability~\cite{zerouali2018formal}. A pivotal concept is \textit{Technical Lag}, defined as the time or version difference between a deployed dependency and its latest available release~\cite{zerouali2018formal, decan2018evolution}. Research by Zerouali et al. has shown that technical lag is a primary vector for security vulnerabilities in software supply chains~\cite{zerouali2021techlag}.

We adapt this socio-technical perspective to the YARA ecosystem by introducing \textit{Mean Time to First Rule} (MTTFR), an ecosystem-level latency metric analogous to technical lag. MTTFR captures the delay between a malware family’s \textit{public disclosure} and the first appearance of a corresponding YARA rule in open-source repositories; \textit{\uline{to the best of our knowledge, this provides the first large-scale empirical measure of ecosystem responsiveness in public YARA rule sharing.}}

A YARA rule is fundamentally an encoded unit of Threat Intelligence (TI). Research on the TI lifecycle emphasizes that Indicators of Compromise (IoCs) suffer from rapid temporal decay~\cite{baranovskyi2023ioc}. Empirical evaluations of open-source TI feeds reveal significant latency, with indicators often being active for weeks before appearing in public feeds~\cite{griffioen2020quality}. Furthermore, the \textit{Pyramid of Pain} model suggests that atomic indicators (IPs, Hashes) typically targeted by simple rules have short operational lifespans compared to behavioral TTPs~\cite{bianco2013pyramid}.
\textit{\uline{Prior YARA research often ignores this temporal dimension. Our analysis quantifies the extent of rule staleness and delayed update propagation (Rule Rot), exposing how the ecosystem acts as an append-only log of decayed indicators rather than a curated intelligence repository.}}
Beyond individual signatures, threat data interoperability remains a major challenge. The STIX \cite{stix_intro} framework standardizes threat information through constructs like indicators and TTPs , enabling automated defensive actions at machine speed. To address data disparities, Andrian et al. \cite{7886562} modeled the STIX architecture using category theory, employing the Functorial Query Language (FQL) for data migration and ologs for machine-compatible knowledge representation. While these works focus on standardized data exchange, our study evaluates the downstream operational reliability of the detection logic itself.

The emerging paradigm of \textit{Detection as Code} (DaC) argues that detection logic should be managed with the same rigor as software code, including version control, CI/CD, and automated testing~\cite{bailey2022detection}. This parallels the evolution of Infrastructure as Code (IaC), where researchers have identified "Security Smells"—recurring coding patterns in scripts (e.g., Ansible, Terraform) that indicate weaknesses~\cite{rahman2019security, kumara2021iac}. Just as IaC smells (e.g., hard-coded secrets) predict deployment failures, we posit that "Detection Smells" (e.g., over-specific metadata, lack of logic modularity) predict operational failure.
\textit{\uline{Our work bridges this gap by applying the rigorous "Code Smell" taxonomy from software engineering to the YARA ecosystem, moving beyond syntax checking to evaluate the structural and semantic quality of detection logic.}}

\section{Methodology}
\label{sec:method}
\begin{figure*}
    \centering
    \includegraphics[width=1\linewidth]{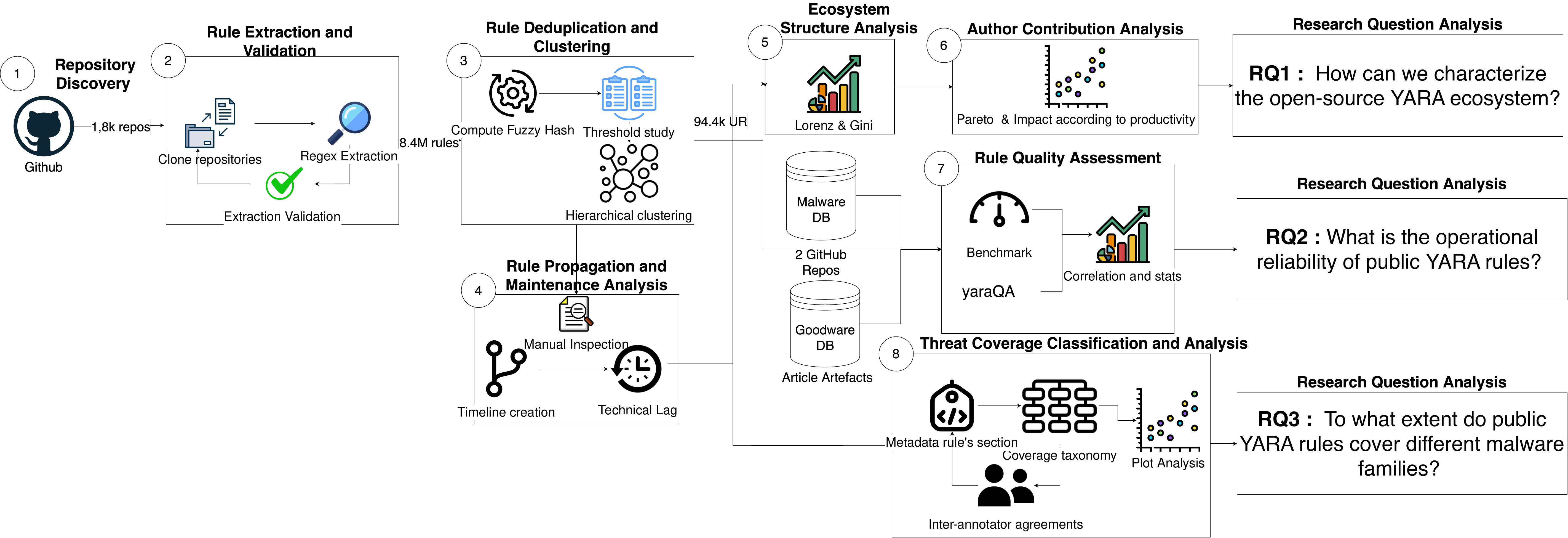}
    \caption{Overview of our empirical study}
    \label{fig:processdiagram}
\end{figure*}
To characterize the open-source YARA ecosystem and address RQ1–RQ3, we developed an eight-stage empirical pipeline integrating repository mining, rule extraction, clustering, and quality assessment. Figure~\ref{fig:processdiagram} illustrates this approach, which combines quantitative and qualitative analyses to capture the ecosystem’s structural and temporal dynamics, and the operational reliability of shared rules. Using a snapshot of 1,853 GitHub repositories containing YARA rules, we analyze their complete histories to uncover large-scale patterns of rule creation, reuse, and evolution. The pipeline proceeds through eight stages: (1) repository discovery and cloning to identify and retrieve YARA projects; (2) rule extraction and validation to parse files, verify syntax, and normalize metadata; (3) rule deduplication and clustering using fuzzy hashing to detect redundancy and shared logic; (4) rule propagation and maintenance analysis to study update frequency and diffusion dynamics through commit histories and technical-lag measurement (RQ1); (5) ecosystem structure analysis to quantify repository-level concentration and redundancy (RQ1); (6) author contribution analysis to measure creator-level concentration and dominance (RQ1); (7) rule quality evaluation through static and dynamic benchmarking (RQ2); and (8) threat coverage classification to map rules to malware families and identify coverage gaps (RQ3). This data-driven methodology provides the foundation for our large-scale characterization of how YARA rules are created, shared, and maintained within the open-source ecosystem.

\subsection{Repository Discovery}
\label{sec:discovery}
To build a dataset representing the open-source YARA ecosystem, we queried the GitHub REST API for repositories labeled \texttt{"Language: YARA"}. This heuristic uses GitHub’s built-in language classifier to identify repositories containing YARA code. The query retrieved 1,853 repositories (snapshot taken on 2025-07-27), which were fully cloned to capture their commit histories and rule files. 

\subsection{Rule Extraction and Validation}
\label{sec:extract}
To extract YARA rules from our repository collection, we implemented a two-phase approach combining targeted file selection with precise pattern matching.
First, we cloned all repositories containing potential YARA files, including specialized extensions (\texttt{.yar}, \texttt{.yara}, \texttt{.rule}, \texttt{.rules}, \texttt{.sig}, \texttt{.yarasig}) and generic text formats (\texttt{.txt}, \texttt{.conf}), as well as extensionless text files. Binary and unrelated artifacts were explicitly excluded to reduce noise and improve efficiency. We opted for an inclusive selection strategy, retrieving all 1,853 repositories to capture the ecosystem in its entirety, including experimental and personal projects. We did not filter forks to analyze propagation (see RQ1), but we explicitly excluded non-YARA binary artifacts during file parsing.
For rule extraction, we developed a regex-based parser to detect the canonical YARA rule declaration:
\begin{verbatim}
^\s*(?:private|global)?\s*rule\s+([A-Za-z0-9_]+)\b
\end{verbatim}
Once a rule header is identified, the parser accumulates content until matching braces, which clearly delimit rule boundaries. This method offers an efficient, scalable alternative to full parsing while maintaining accuracy comparable to existing tools.
To ensure correctness, we implemented a validation protocol. We utilized \textit{Plyara}~\cite{Plyara_GitHub}—a standard Python library based on the PLY (Python Lex-Yacc) implementation of an LALR parser—as our ground truth.
Given the prohibitive computational cost of formally parsing the entire corpus (8.4 million files), we applied \textit{Plyara} for cross-validation on a statistically significant random subset of the dataset (10\%, approx. 840k rules). We achieved 100\% agreement on a random 10\% subset, where the validation was not limited to counts but included the exact string contents, identifier names, and the full semantic block of the condition. This operational definition of "agreement" ensures that the logic utilized for downstream clustering and quality assessment is identical to that of a formal grammar-based extraction. This process confirms that, within the scope of the extracted attributes (imports, rule names, string counts), our optimized regex approach is empirically equivalent to the formal LALR parser while remaining computationally efficient for large-scale analysis.

\subsection{Rule Deduplication and Clustering}
\label{sec:cluster}
YARA rules authored by different contributors often contain minor, non-semantic differences—such as formatting changes, field reordering, variable renaming, or comment edits—that do not alter their detection logic but fragment identical logic across multiple repositories. This fragmentation obscures patterns of rule reuse, collaboration, and evolution across the ecosystem.

To enable large-scale analysis of redundancy, origin, and propagation, we first need a representation that captures when different rules implement the same underlying logic, even if they are not textually identical. We used an approach similar as J. Kornblum \cite{KORNBLUM200691}, we therefore apply fuzzy hashing~\cite{plagiarismDetection2023} to generate similarity-preserving fingerprints for each rule. Identical rules yield identical hashes, while near-duplicates produce highly similar ones. Specifically, we use \textit{ssdeep}, a widely adopted implementation of context-triggered piecewise hashing (CTPH), which computes similarity scores between strings on a 0–100 scale. This enables detection of near-duplicate rules and their aggregation into clusters, where each \textbf{cluster} represents a set of mutually similar rules sharing the same \textbf{unique rule logic (UR)}.

To verify the suitability of \textit{ssdeep} for YARA rule comparison, we constructed a controlled benchmark dataset based on the curated ruleset published by Florian Roth~\cite{roth2023yaraforge}, which contains non-duplicate “original’’ rules. From this baseline, we introduced two categories of modifications: \textit{variants} and \textit{mirages}. Variants represent true near-duplicates, created through non-semantic edits such as variable renaming, metadata changes, or minor logical adjustments. In contrast, mirages represent false positives—syntactically similar but logically distinct rules—generated by mixing or altering strings and conditions from unrelated rules. Each rule in the benchmark was labeled as \textbf{original}, \textbf{variant}, or \textbf{mirage}, enabling quantitative evaluation of fuzzy-hashing performance.

We computed \textit{ssdeep} hashes for each rule and evaluated multiple similarity thresholds under two configurations: (1) hashing the entire rule content (including metadata) and (2) a \textit{logic-only} mode that excludes metadata to avoid author-style bias. For each threshold, we recorded True Positives (TP)—variants correctly matched to originals—and False Positives (FP)—unrelated rules incorrectly matched. The optimal threshold was selected to maximize the F1-score, balancing sensitivity and specificity.

Both configurations converged on an optimal similarity threshold of 65\%, achieving an F1 of 0.77 on our challenging benchmark. Despite the dataset’s adversarial design, this result shows that \textit{ssdeep}, when properly tuned, remains robust and effective for detecting near-duplicate YARA rules under realistic conditions.

Using the optimized threshold, we computed \textit{ssdeep} hashes for all extracted rules and applied hierarchical clustering~\cite{Barman2016}. To handle the significant computational complexity of performing pairwise comparisons across 8.4 million rules ($O(n^2)$), we implemented a \textbf{blocking strategy}. We first canonicalized the rule text by stripping comments and normalizing whitespace to ensure that non-semantic formatting did not influence the fuzzy hashes. 

We then partitioned the dataset into blocks based on the \textit{ssdeep} signature's block key (the first 24 characters), effectively grouping rules with a high prior probability of similarity. Within each block, pairwise distances were computed as $d = 1 - (\textit{similarity} / 100)$, where \(\textit{similarity} \in [0, 100]\) is the percentage score returned by the fuzzy-hash comparison. We applied hierarchical clustering using the \textbf{average linkage method} and a distance threshold of 0.35, corresponding to our 65\% similarity requirement. 

This clustering step produced two key artifacts: (i) a \emph{cluster inventory} consisting of 94.4k unique rule logics and (ii) a \emph{repository--cluster incidence matrix}. This matrix serves as the structural backbone for subsequent analyses, supporting the computation of repository-level concentration and redundancy (RQ1), author-level contribution distributions (RQ2), and temporal propagation dynamics (RQ3).

\subsection{Rule Propagation and Maintenance Analysis}
\label{sec:propagation}

To support a deeper temporal analysis of the ecosystem (RQ1), we traced each extracted YARA rule back to the commits that introduced or modified it. This enables us to study rule propagation across repositories and to characterize maintenance and update diffusion over time.

A manual review of the top 10\% of repositories (by unique rule logic count) and the top/bottom 1\% of clusters (by cardinality) to interpret structural patterns. This qualitative step contextualizes maintenance trends by identifying repository archetypes (e.g., rule-dump hubs vs.\ curated feeds). All assessments were independently double-annotated by two researchers for reliability. We achieved a Cohen’s Kappa agreement of 0.92 (strong agreement). Disagreements were explicitly discussed and resolved to refine the classification.
Specifically, we distinguish between \textbf{source repositories}, which introduce new rules, and \textbf{mirror repositories}, which primarily replicate existing ones. We make this distinction by computing the \textbf{first-publisher ratio} for each repository, defined as:
\[
\text{first-publisher ratio} = 
\frac{\#\text{clusters first appearing in the repository}}
{\#\text{clusters contained in the repository}}.
\]
A ratio close to 1 indicates a repository that primarily creates original rules (a source), while a ratio near 0 denotes one that mainly copies existing rules (a mirror).  
For example, if a repository contains 200 rule clusters, 160 of which first appeared there, its first-publisher ratio is \(0.8\), classifying it as a source.  
Our empirical distribution of this ratio is strongly bimodal, confirming that most repositories fall distinctly into one of these two categories.  
We thus classify repositories with a ratio $\geq 0.80$ as \textbf{sources} and those with $\leq 0.20$ as \textbf{mirrors}, allowing for mixed cases that import some external rules while remaining predominantly original or derivative.

To quantify the \textbf{propagation delay} of rules, we measure their \textbf{technical lag}~\cite{TFModuleUpdates2025}.  
For a given cluster \(c\) and mirror repository \(r\), we compute:
\[
\text{lag}(c,r) = 
\min_{t \in U(r,c)} \bigl[t - t_s(c)\bigr],
\]
where \(t_s(c)\) is the timestamp of the cluster’s first appearance in its source repository, and \(U(r,c)\) is the set of commit times in repository \(r\) that modify files containing cluster \(c\).  
This metric captures the delay between a rule’s creation and its subsequent adoption by mirrors, enabling analysis of diffusion speed and maintenance synchrony both globally and per repository.

A key methodological consideration is that Git diffs operate at the \emph{file} level rather than per individual rule.  
When multiple rules coexist in the same file, update tracking may overestimate propagation delay.  
To address this, we (a) report the average number of rules per file to assess granularity, and (b) recompute lag metrics using only single-rule files as a robustness check.  
This two-pronged validation ensures our findings are not biased by file-level approximations.

\subsection{Ecosystem Structure Analysis}
\label{sec:structure}
To understand how YARA rules are produced and reused across the open-source ecosystem, we analyze the distribution of rule clusters across repositories to assess whether rule development is widely shared or concentrated in a few dominant hubs. We focus on two key structural properties: \textbf{concentration} and \textbf{redundancy}. Concentration captures how strongly rule creation and sharing are centralized, while redundancy measures how often identical or near-identical rules recur across repositories. Both are quantified using scale-independent metrics that reveal inequality and duplication patterns within the ecosystem.

We employ three complementary analyses to capture these dimensions.  
To measure \textbf{concentration}, we analyze the distribution of \textbf{clusters per repository} (the number of unique clusters in a repository, or \textbf{repository cluster count}) using Lorenz curves and Gini coefficients ($G_{contribution}$)~\cite{OlgaGini}. The Lorenz curve plots the cumulative share of clusters (y-axis) against the cumulative share of repositories sorted by cluster counts (x-axis), where the $45^\circ$ line represents perfect equality. The Gini coefficient, calculated as
\[
G = 1 - 2 \int_0^1 L(p)\, dp,
\]
with $L(p)$ denoting the Lorenz curve, quantifies inequality in this distribution. A value of $G=0$ denotes perfect equality, while $G=1$ indicates maximal inequality. Higher $G$ values thus indicate stronger \emph{centralization}. We report 95\% bootstrap confidence intervals based on $10^4$ resamples.

To measure \textbf{redundancy}, we compute the \textbf{cluster cardinality} (the number of distinct repositories containing a given UR) and its Gini index ($G_{popularity}$), capturing duplication patterns and identifying highly replicated rules.

\subsection{Author Contribution Analysis}
To attribute authorship, we identify the original creator of each unique rule cluster by tracing the earliest commit introducing that cluster across all repositories. Author metadata (name and email) are extracted from Git histories and normalized to resolve aliases, allowing consistent attribution of each cluster to a single author.

We then analyze the distribution of author contributions. An author's \textbf{productivity} is defined as the number of unique rule clusters they created, and their \textbf{impact} as the cumulative number of appearances of those clusters across repositories. We quantify concentration of influence through Pareto distribution analysis~\cite{Veeger2025Pareto} and examine the relationship between productivity and impact, as shown in Figure~\ref{fig:prodimpact}.

Finally, we qualitatively examine the most prolific authors (top 1\%) to contextualize their roles in the ecosystem—such as researchers, security firms, or community hubs. This analysis captures creator centralization and complements our repository-level findings by revealing whether concentration arises from project structure or dominance of a few key contributors.

\subsection{Rule Quality Assessment}
\label{sec:quality}

After examining how YARA rules are created, shared, and propagated, we assess their \textbf{quality}—whether they are well-constructed and effective at detecting real threats. Evaluating rule quality is key to understanding the practical reliability of the open-source YARA ecosystem, as even widely shared or well-maintained rules may be syntactically flawed, logically inconsistent, or overly broad. Addressing RQ2, we assess both the \textbf{technical soundness} and \textbf{practical effectiveness} of rules through static analysis and operational benchmarking.

To assess \textbf{technical soundness}, we apply \textit{yaraQA}~\cite{neo23x0_yaraqa}, which assigns each rule a base score of 100 points and deducts penalties according to issue severity (1–70 points). This static analysis examines metadata completeness (author, description, reference), logical consistency of conditions, pattern efficiency, and the absence of problematic constructs such as unsatisfiable conditions or overly broad patterns. Scores are aggregated at the unique rule logic (UR) level to enable repository-level and ecosystem-wide comparisons.

To evaluate \textbf{practical effectiveness}, we benchmark rule performance against real samples. Each rule is tested against (a) a \textit{goodware corpus} comprising both core system files from fresh Windows installations and popular applications scraped from legitimate software portals, as originally collected by Fasci et al ~\cite{Fasci2023Disarming} for false-positive measurement and (b) malware samples from \textit{TheZoo}~\cite{thezoo_github_2025} and honeypots for true-positive detection.
To avoid bias, we consider only rules producing at least one positive match (TP or FP), since unmatched rules may target unseen threats rather than being faulty. From these results, we compute standard detection metrics—precision, recall, and F1—to quantify accuracy.

Finally, we measure parsing and scan times and analyze the correlation (Spearman’s $\rho$) between false-positive rates and performance overhead to explore the trade-off between detection accuracy and computational efficiency.
We utilized Spearman’s rank correlation ($\rho$) instead of Pearson’s because the distributions of parsing times and false positive rates were non-normal (confirmed via Shapiro-Wilk tests, $p\le0.05$).
This combined approach provides both technical and empirical evidence of rule quality across the open-source YARA ecosystem. All performance benchmarks (scanning speed and parsing time) were executed on the following environent: Intel Core Ultra 7 265K CPU, 64GB RAM, running Windows 11 Pro. To minimize noise, each measurement was repeated 10 times and averaged.

\subsection{Threat Coverage Classification and Analysis}
\label{sec:coverage}
In section~\ref{sec:quality}, we assessed the effectiveness of individual YARA rules. We now extend this analysis to the ecosystem level by examining \textbf{coverage}—whether open-source rules collectively address the breadth of real-world threats. Addressing RQ3, we measure the \textbf{scope and evolution of threat coverage} to determine if the right malware families are represented and where blind spots persist.

To measure coverage, we map each unique rule logic (UR) to a \emph{malware family} and compute two complementary metrics:
\[
C_f = \frac{\#\text{UR labeled } f}{\#\text{all UR}}, \qquad
C_f^{\text{repo}} = \frac{\#\text{repositories with }\geq 1 \text{ UR for } f}{\#\text{all repositories}}
\]
Here, $C_f$ (\textit{Rule-Level Coverage}) quantifies the \textbf{depth} of coverage—how many unique logics target family $f$—while $C_f^{\text{repo}}$ (\textit{Repository-Level Coverage}) measures the \textbf{breadth} of adoption across repositories. Together, these metrics capture whether detection effort is concentrated on a few threat types or evenly distributed across the malware spectrum.

We implement a semi-automated classification framework combining metadata parsing with a curated taxonomy. The taxonomy spans 17 primary malware categories and 196 alias terms, covering families such as \textit{Ransomware, RAT, Stealer, Backdoor, Cryptominer}, and others. Each category includes representative aliases (e.g., “ransom”, “ryuk”, “lockbit” for \textit{Ransomware}) used as triggers. This taxonomy is derived from two previous works \cite{MalwareDetectionML,taxbased}.

For each rule, we extract key elements using regular expressions applied to:  
(1) the rule name, (2) optional family headers, and (3) the metadata block:

\begin{verbatim}
rule_name_re = re.search(r"rule\s+([^\s:{]+)", text)
header_re = re.search(r"rule\s+
[^\s{]+(?:\s*:\s*([^{\n]+))?", text)
meta_block_re = re.compile(r"meta:\s*
(.*?)\s*(?:strings:|condition:)",re.S | re.I)
\end{verbatim}

Tokens obtained from these components are normalized and matched against alias terms to assign each UR to a malware family. All alias definitions and parsing scripts are available in our replication package.

We analyze temporal dynamics by recording the first appearance date of each UR and aggregating by family. Results are visualized using \textit{beeswarm plots} (Fig.~\ref{fig:beeswarm}), where point size indicates cluster cardinality (number of rules sharing the same logic). This highlights the timing and relative popularity of coverage across families.

To ensure reliability, two authors independently annotated a random sample of 10\% of the dataset. We achieved a Cohen’s Kappa agreement of 0.82 (strong agreement). Disagreements were explicitly discussed and resolved to refine the classification taxonomy. This validation confirms that observed coverage imbalances reflect real detection gaps rather than classification noise.
To quantify the ecosystem's reactivity, we introduce the \textit{Mean Time to First Rule} (MTTFR) inspired from the Mean Time to Repair metric in incident management. This metric measures the delay between the official public disclosure of a malware family (Ground Truth) and the appearance of its first detection rule in the open-source ecosystem. 
Unlike previous studies that rely on file creation timestamps which are often misleading due to the practice of appending new rules to existing files. We developed a specific heuristic. Our extraction pipeline parses commit messages (e.g., "Added WannaCry detection") and identifies family-specific filenames (e.g., \texttt{wannacry.yar}) to pinpoint the exact moment of \textit{detection intent}.
We applied the MTTFR metric to a curated set of 14 malware families, strategically selected to contrast two distinct threat profiles. This comparative approach allows us to determine whether the open-source community prioritizes reacting to high-impact events or proactively detecting initial access vectors.

Overall, this analysis quantifies how comprehensively public YARA rules cover known malware categories and where detection efforts remain limited.

\section{Results}
\label{sec:results}

This section presents the findings of our empirical analysis. We characterize the ecosystem's socio-technical dynamics (RQ1), evaluate operational reliability (RQ2), and assess semantic alignment with the threat landscape (RQ3).

\subsection{RQ1: Ecosystem Dynamics and Maintenance}
\label{sec:results_rq1}
To answer RQ1, we characterized the ecosystem’s socio-technical dynamics by mapping its structure, authorship, and temporal evolution. Our results reveal that the open-source YARA landscape functions primarily as a \textbf{centralized and static archival network} rather than a dynamic software supply chain. While the ecosystem appears vast in scale—containing over \textbf{8.4 million rule occurrences} it is operationally unreliable. It is defined by three converging phenomena: extreme \textbf{structural redundancy} (where 89\% of content is duplicated), \textbf{centraliazed control} (where 1.3\% of authors drive 80\% of adoption), and \textbf{temporal stagnation} (with a median technical lag of 4.2 years). The following subsections detail these structural, social, and temporal dimensions.

\subsubsection{RQ1.1 Structure: A Redundant and Fragmented Landscape}
Our analysis of 1,853 repositories reveals an ecosystem characterized by extreme redundancy and structural polarization. From a raw dataset of \textbf{8.4 million rule occurrences}, our fuzzy clustering identified only \textbf{94.4k Unique Rule logics (URs)}. This represents a massive compression ratio, confirming that the vast majority of the ecosystem consists of duplicated content.

\textbf{Finding 1: High Redundancy and Ad-hoc Propagation.}
Unlike mature software ecosystems (e.g., npm or PyPI) that manage reuse through explicit dependency declarations~\cite{decan2018evolution}, the YARA landscape is dominated by static file duplication. We observe that rules are physically replicated across repositories rather than referenced. This mechanism results in a fragmented maintenance landscape: when a rule is updated in a source repository, the replicated instances elsewhere remain static, effectively detaching them from upstream improvements.

\textbf{Finding 2: A Bimodal Repository Distribution.}
We analyzed the \textit{First-Publisher Ratio} to characterize how repositories contribute to the ecosystem. The distribution is strongly bimodal, allowing us to categorize repositories into two distinct archetypes:
\begin{itemize}
    \item \textbf{Mirrors (Aggregators):} The majority of repositories function as archival sinks. They ingest thousands of rules with near-zero originality, prioritizing quantity over curation.
    \item \textbf{Sources (Innovators):} A smaller subset generates new logic. However, we observe a critical "Innovation Isolation": \textbf{26.8\% of unique rules} appear in only a single repository (Singletons). These contributions, often from small or academic projects, fail to propagate to the wider community.
\end{itemize}

\subsubsection{RQ1.2 Author Influence and Centralization}
The distribution of influence is heavily skewed ($G=0.761$), indicating a concentration of contributions ecosystem.

\begin{figure}[t]
    \centering
    \includegraphics[width=1\linewidth]{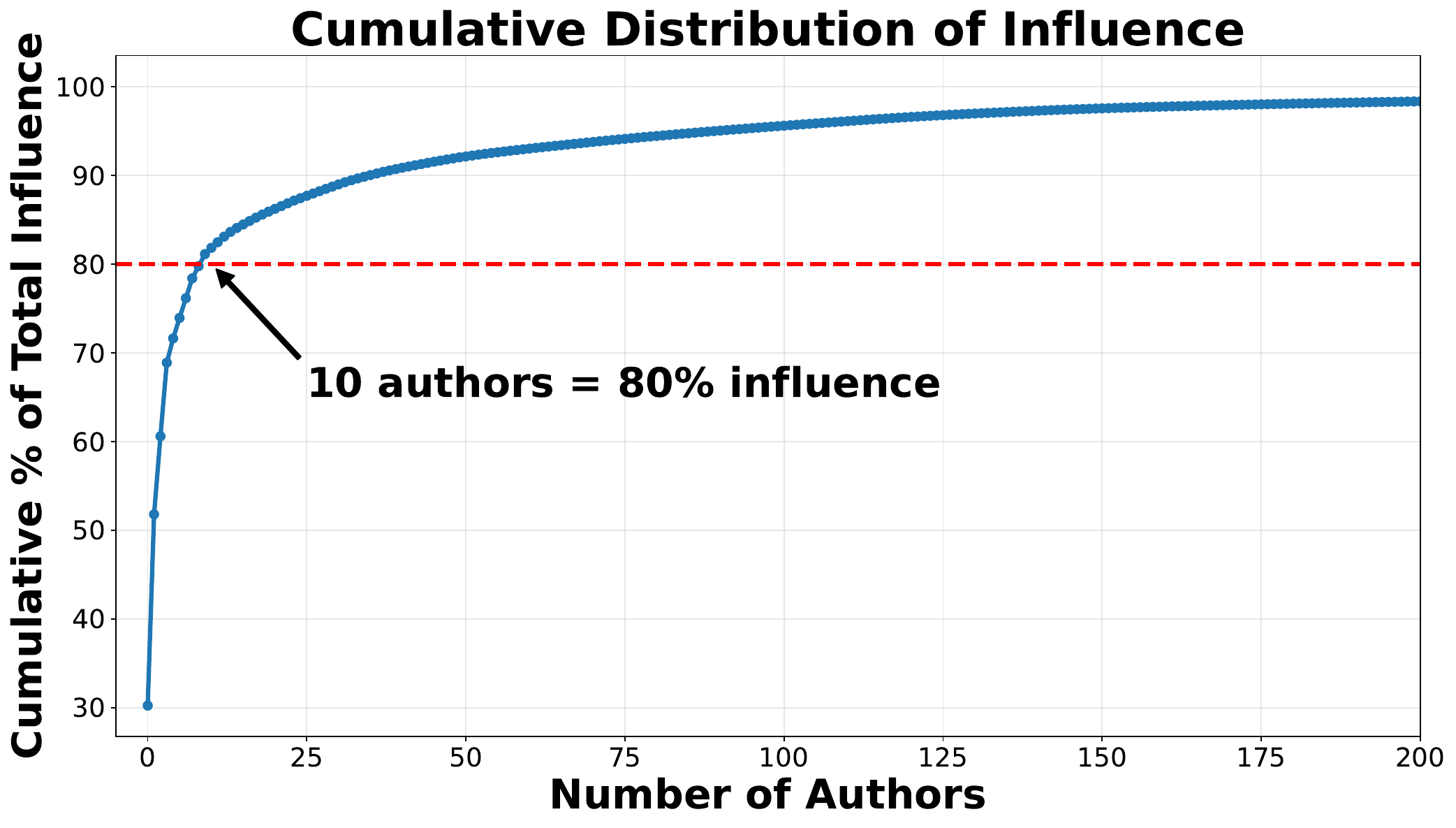}
    \caption{Pareto curve of author influence: A core group of 10 authors drives 80\% of all rule adoption.}
    \label{fig:paretochart}
\end{figure}

\textbf{Finding 3: Centralized structure}
As shown in Figure~\ref{fig:paretochart}, the ecosystem is driven by a core group of just \textbf{10 authors} (approx. 1.3\% of the identified population) accounts for \textbf{80\% of the total ecosystem impact}, measured by cumulative rule adoption.

\begin{figure}[t]
    \centering
    \includegraphics[width=1\linewidth]{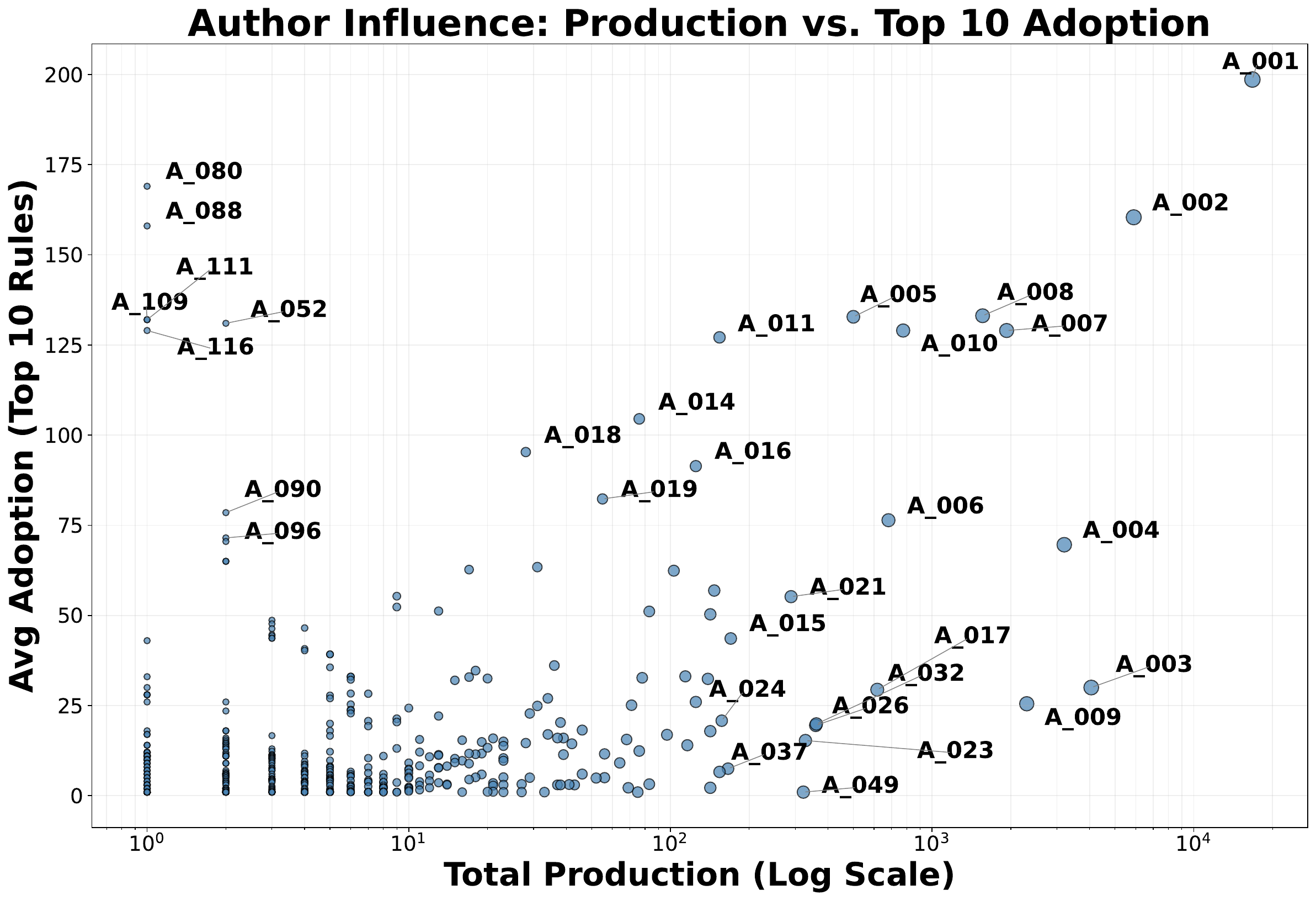}
    \caption{Influence Map: Comparing author production volume vs. peak impact reveals distinct archetypes (Specialists vs. Mass Producers).}
    \label{fig:prodimpact}
\end{figure}

\textbf{Finding 4: Distinct Contributor Archetypes.}
By mapping authors based on Production Volume versus "Peak Impact" (average adoption of their top 10 rules), Figure~\ref{fig:prodimpact} reveals two primary behaviors. \textit{Mass Producers} (e.g., author \texttt{A\_001}) contribute immense volumes of signatures, serving as "bulk suppliers." Conversely, \textit{High-Efficiency Specialists} produce fewer rules but achieve exceptional saturation, effectively acting as the "standard library" of the ecosystem.

\subsubsection{RQ1.3 Evolution: The Static Supply Chain}
Our temporal analysis suggests the ecosystem functions primarily as a static archive rather than an active supply chain.

\textbf{Finding 5: High Latency and Stagnation.}
Activity levels are low, with only \textbf{30.7\%} of repositories showing a commit in the last year (median time since last update: \textbf{782 days}). To quantify the freshness of threat intelligence, we measured \textbf{Technical Lag}—the delay between a rule's creation and its adoption in mirror repositories. The median lag is \textbf{1,525 days} ($\approx$ 4.2 years), with a 95th percentile exceeding 12 years.

We observe a significant divergence in maintenance models. While curated vendor feeds (e.g., \texttt{reversinglabs}) maintain a \textbf{0-day lag}, massive aggregators like \texttt{ContainSAGES} (hosting $>$2.4M rules) exhibit an average lag of \textbf{3,021 days}. This confirms that while rule volume is high, the "freshness" of the available intelligence in the wider ecosystem is severely degraded.

\begin{tcolorbox}[colback=blue!5!white, colframe=black!75!black, title=Summary of Findings for RQ1]
The YARA ecosystem is defined by structural concentration and temporal stagnation. It relies on a \textbf{core of 10 authors} who drive 80\% of rule adoption. However, due to the lack of dynamic dependency management, the median \textbf{Technical Lag reaches 4.2 years}, rendering a large portion of the shared ecosystem historically significant but operationally outdated.
\end{tcolorbox}

\subsection{RQ2: Operational Reliability and Quality}
\label{sec:results_rq2}

\begin{figure}[t]
    \centering
    \includegraphics[width=1\linewidth]{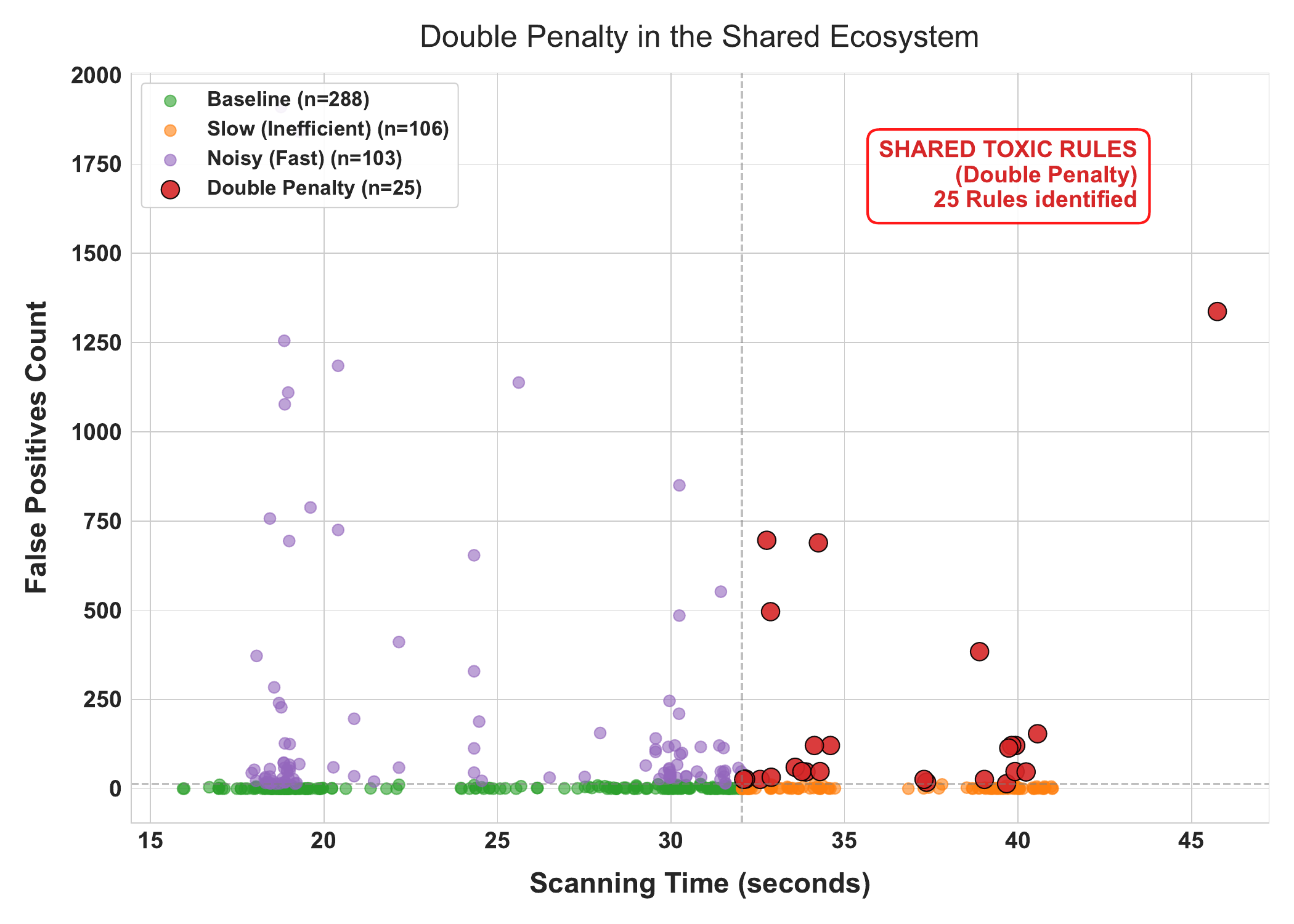}
    \caption{Performance vs. Reliability in the Shared Ecosystem: The "Double Penalty" zone (red) highlights rules that are both computationally expensive and operationally noisy.}
    \label{fig:double_penalty}
\end{figure}

Having established the ecosystem's structural stagnation, we assessed the operational quality of the 14,595 rules constituting the "Shared Ecosystem" (rules adopted by at least 3 repositories).

\textbf{Finding 6: High Static Validity vs. Operational Noise.}
Statistically, the ecosystem appears robust: the mean static quality score is \textbf{99.45/100}, and syntax errors are virtually absent. However, this metric is misleading. Our dynamic benchmarking reveals a significant discrepancy between static validity and operational utility. A substantial portion of valid rules are either functionally dormant or excessively noisy when executed against real datasets.

\textbf{Finding 7: Absence of Cost-Quality Trade-off.}
Intuitively, deeper (and slower) inspection should yield higher precision. However, our analysis refutes this assumption for the current ecosystem. We observed a weak negative correlation between scan time and false positive rates ($\rho = -0.154$).
This indicates that computational cost is rarely an investment in deeper inspection but rather a symptom of poor rule design.

\textbf{Finding 8: The "Double Penalty" Phenomenon.}
We identified three distinct operational archetypes (Figure~\ref{fig:double_penalty}), confirming and extending prior work on YARA performance~\cite{regeciova2023yara}:

\begin{itemize}
    \item \textbf{The "Noisy-Fast" Rules:} Efficiency does not imply quality. Rules like \texttt{check\_RaiseException\_iat} execute rapidly (18s) because they check for standard API imports. However, their lack of specificity results in \textbf{372 False Positives}, rendering them functionally indistinguishable from noise.
    
    \item \textbf{The "Double Penalty" (Toxic Rules):} We identified a cluster of rules that fail on both performance and precision axes. The rule \texttt{contentis\_base64} is a paradigmatic example. As predicted by Regéciová et al.~\cite{regeciova2023yara}, its reliance on unoptimized regular expressions (\texttt{/([A-Za-z0-9+\/]{4})\{3,\}/}) causes extreme scan times ($45.7s$). Crucially, our data adds an operational dimension to this mechanical finding: this technical inefficiency is coupled with \textbf{1,337 False Positives}. Adopters of such rules incur a high computational cost for alerts that are nearly 100\% false.
    
    \item \textbf{The High-Value Minority:} Conversely, rare examples like \texttt{SierraBravo\_Two} demonstrate that efficiency is possible. By combining specific bytecode signatures with structural constraints (PE section targeting), it achieves high true-positive rates with zero noise and fast execution (17.5s).
\end{itemize}

The persistence of the "Double Penalty" rules (red region in Figure~\ref{fig:double_penalty}) within the shared ecosystem—despite their demonstrable inefficiency—suggests a lack of evolutionary pressure. Ineffective logic is preserved and replicated across repositories alongside high-quality rules, creating significant technical debt for adopters.

\begin{tcolorbox}[colback=blue!5!white, colframe=black!75!black, title=Summary of Findings for RQ2]
Operational reliability is decoupled from static validity. We refute the existence of a rational cost/quality trade-off: expensive rules are often the least precise. This "Double Penalty" phenomenon means adopters frequently inherit shared rules that simultaneously degrade pipeline performance and flood SOCs with false positives.
\end{tcolorbox}

\subsection{RQ3: Semantic Threat Coverage}
\label{sec:results_rq3}

\begin{figure*}[t]
    \centering
    \includegraphics[width=.8\linewidth]{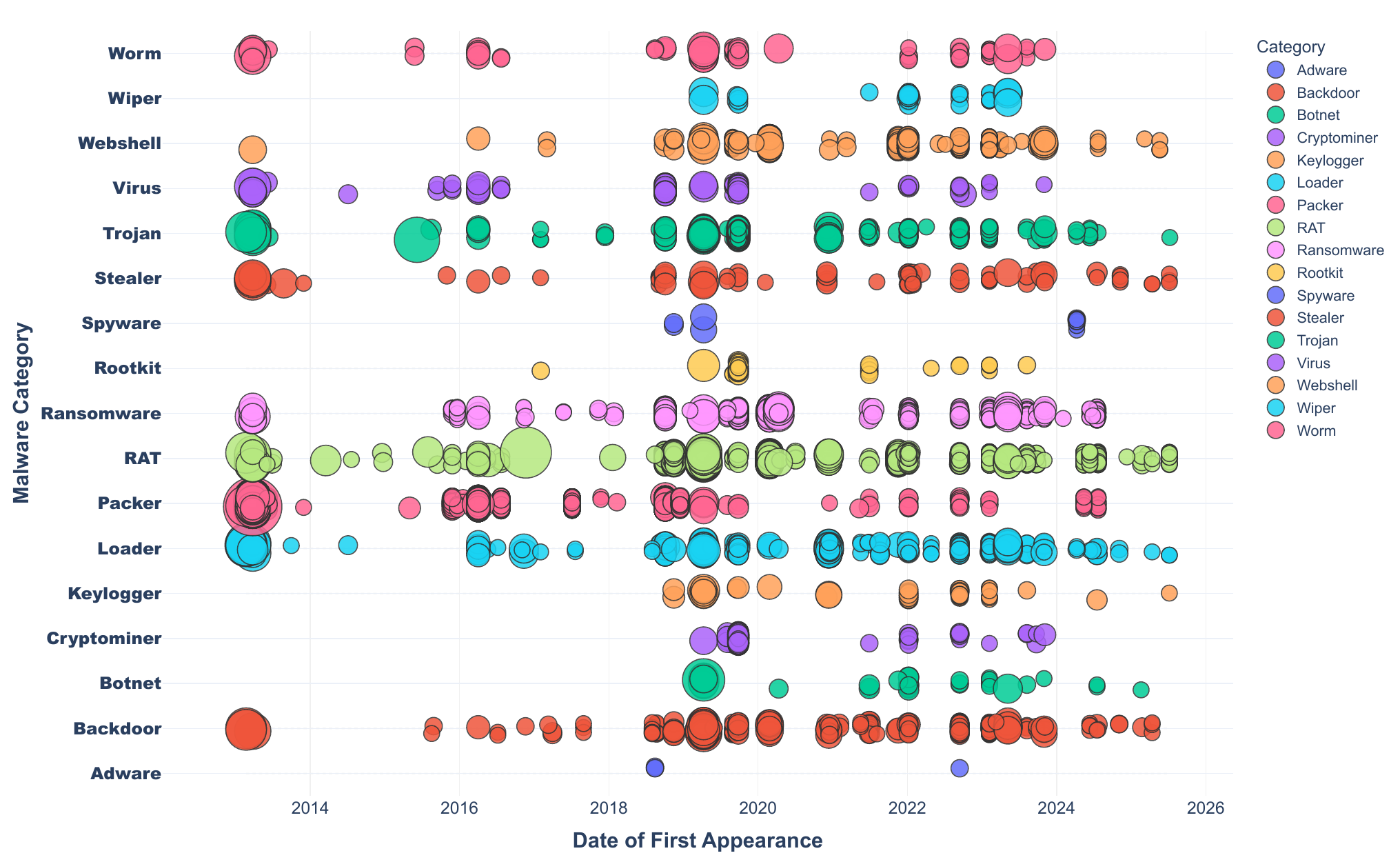}
    \caption{Temporal emergence of YARA rules by malware family. Circle size represents cluster cardinality (reuse), showing a heavy bias towards Ransomware and Trojans.}
    \label{fig:beeswarm}
\end{figure*}

Finally, we mapped the ecosystem's output against the malware landscape to identify coverage biases and reaction times. Figure~\ref{fig:beeswarm} illustrates the temporal emergence and accumulation of rules across major malware families.

\textbf{Finding 9: Bias Towards "High-Visibility" Legacy Threats.}
The ecosystem is historically dominated by \texttt{Trojan} and \texttt{Ransomware} families. As shown in Figure~\ref{fig:beeswarm}, these categories exhibit early appearance (2012–2014) and massive reuse (large cluster cardinality). This concentration correlates with waves of highly publicized global incidents (e.g., WannaCry, NotPetya), suggesting that community effort is primarily driven by the visibility of the impact rather than the technical sophistication of the threat.

\textbf{Finding 10: The Precursor Gap (Initial Access).}
In contrast, distinct "Precursor" families—specifically \texttt{Loaders}, \texttt{Stealers}, and \texttt{Droppers}—remain significantly underrepresented. Despite their foundational role in modern intrusion chains (often serving as the delivery mechanism for ransomware), they show limited cluster cardinality and late adoption. Loaders such as \textit{BazarLoader} and \textit{QakBot} only appear in significant numbers after 2018. Referencing Bianco's "Pyramid of Pain"~\cite{bianco2013pyramid}, our data indicates that the open-source ecosystem focuses heavily on static, high-level artifacts (the wide base of the pyramid) while struggling to address the evolving TTPs of initial access brokers.

\begin{figure}[t]
    \centering
    \includegraphics[width=1\linewidth]{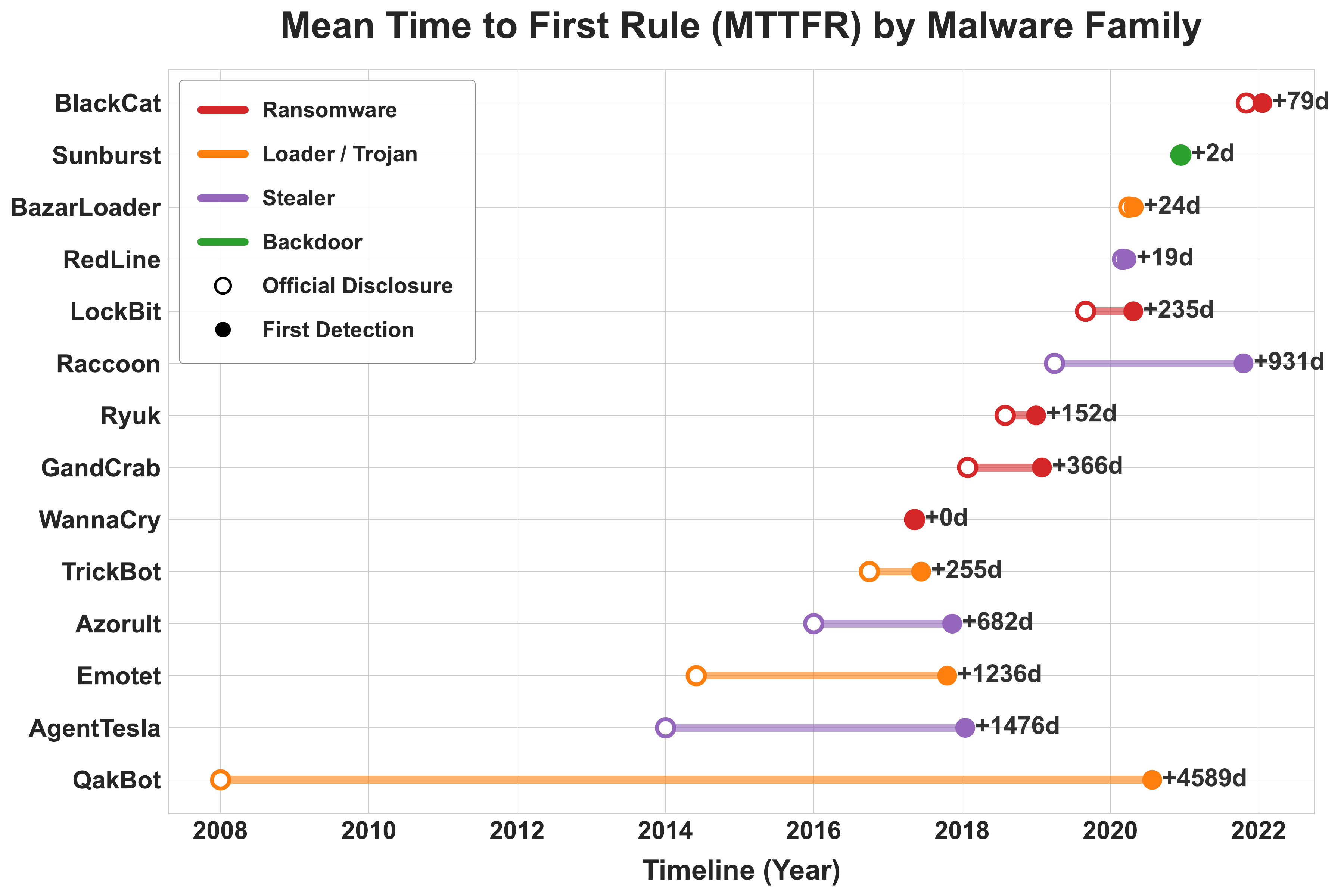}
    \caption{Mean Time to First Rule (MTTFR) for 14 specific malware families.}
    \label{fig:MTTFR}
\end{figure}

\textbf{Finding 11: Reactive Latency vs. Silent Dwell Time.}
To quantify the speed of community mobilization, we calculated the \textit{MTTFR} for 14 representative threats (Figure~\ref{fig:MTTFR}). The results reveal a stark temporal disparity driven by the threat profile. High-profile, disruptive events trigger near-instantaneous community mobilization. For instance, \textit{WannaCry} (Ransomware) shows a lag of \textbf{0 days}, and the highly publicized supply-chain attack \textit{Sunburst} (Backdoor) was covered within \textbf{2 days}. Similarly, recent ransomware strains like \textit{BlackCat} show a relatively short reactive lag (+79 days).

Conversely, stealthy infrastructure threats exhibit extreme latency. \textit{QakBot}, a banking trojan turned loader, existed for \textbf{4,589 days} (over 12 years) before specific detection rules appeared in our public dataset. Similarly, critical threats like \textit{AgentTesla} (+1,476 days) and \textit{Emotet} (+1,236 days) operated largely unchecked by the open-source community for years. This data quantitatively confirms that the ecosystem functions primarily as a \textit{reactive} mechanism: it excels at documenting the aftermath of noisy attacks but fails to proactively secure the initial access vectors that facilitate them. This gap aligns with industry observations~\cite{crowdstrike2025,mtrends2025}, which note that while loaders underpin the majority of MaaS operations, they remain the "blind spot" of public detection engineering.

\begin{tcolorbox}[colback=blue!5!white, colframe=black!75!black, title=Summary of Findings for RQ3]
Coverage is driven by visibility, not operational risk. The ecosystem reacts instantly to high-profile disruptive events (MTTFR $\approx$ 0 days) but exhibits extreme blindness towards "silent" Initial Access vectors. Critical loaders like \textit{QakBot} remained undetected in public repositories for over \textbf{12 years} (4,589 days), highlighting a systemic failure to address the early stages of the kill chain.
\end{tcolorbox}

\section{Implications}
\label{sec:discussion}

Our findings highlight a gap between the ecosystem's scale and its operational utility. By mapping this landscape, we identify critical shifts required for the three pillars of the security community.

\textbf{For Academic Research: From Syntax to Semantics}
Current literature focuses heavily on automated rule generation, often using validity as the sole success metric. Our "Double Penalty" analysis demonstrates that this is insufficient: a rule can be valid yet operationally toxic. Consequently, researchers must pivot to \textbf{"Detection Engineering"} metrics. Future studies should stop treating rules as static text and start evaluating them as executable code, defining specific "Detection Smells"—such as greedy patterns or high-cardinality strings—that predict operational failure. We need new frameworks to quantify the \textit{maintainability} and \textit{robustness} of signatures, akin to Technical Debt metrics in Software Engineering.

\textbf{For Practitioners: The "Raw Data" Paradigm}
For SOC managers and CTI analysts, the ecosystem's structural stagnation proves that public repositories cannot be trusted as "live" intelligence feeds. Practitioners must therefore move from passive collection to active curation. Given the extreme technical lag identified, analysts should adopt a \textbf{"Distrust by Default"} mindset, assuming any public rule is outdated unless proven otherwise. Furthermore, instead of importing from massive aggregator repositories which introduce noise, defenders should trace rules back to their "Source" repositories to ensure they receive upstream updates. Finally, because the community's bias towards Ransomware leaves Initial Access vectors exposed, SOCs must proactively develop internal coverage for Loaders and Stealers, as the open-source ecosystem consistently fails to cover these early-stage threats in a timely manner.

\textbf{For Tool Builders: Profiling over Linting}
The link between performance overhead and false positives indicates that execution cost is a strong proxy for quality. Yet, existing tools are limited to static syntax checking. The tooling ecosystem must therefore evolve from "Linters" to \textbf{"Profilers."} Just as software CI/CD pipelines use unit tests, YARA pipelines need dynamic benchmarking tools that automatically reject rules falling into the "Critical Region" of performance cost. Preventing toxic logic from entering production is fundamentally more effective than filtering alerts downstream.

\section{Threats to Validity} \label{sec:threats}
\textbf{Ecosystem Representativeness.}
Our dataset of 8.4M GitHub rules excludes private, commercial, or governmental rulesets, which may follow different standards. Thus, findings reflect open-source behavior and may not generalize to all YARA deployments. Restricting the study to GitHub may introduce platform-specific bias, as some communities utilize GitLab or self-hosted instances.

\textbf{Internal Process Robustness.}
Pipeline reliability was ensured by cross-validating our regex parser against \textit{Plyara} on a significant 10\% subset, achieving 100\% agreement on extracted attributes. Additionally, the 65\% \textit{ssdeep} clustering threshold was optimized via sensitivity analysis on a benchmark of variants and "mirage" rules (F1-score: 0.77), ensuring robust redundancy and authorship metrics at scale.

\textbf{Construct Validity (Measurement Proxies).}
Classification and temporal metrics rely on proxies. Malware mapping uses metadata-based heuristics; ambiguous data may lead to misclassification. To mitigate Git timestamp distortion from bulk imports, we utilized a temporal heuristic validating "detection intent" via commit semantics. Furthermore, focusing strictly on YARA logic captures actionable detection but excludes the broader standardized exchange context provided by formats like STIX/TAXII or categorical modeling frameworks.

\textbf{Limitations of Quality Assessment.}
Our evaluation is two-faceted. Static analysis (\textit{yaraQA}) confirms syntax but not operational noise or recall. Similarly, our dynamic benchmark was constrained by sample diversity and laboratory conditions. These tests are intended to highlight the gap between static scores and practical utility rather than providing absolute performance measures, reinforcing the need for data-driven curation.

\section{Conclusion}
This paper provides the first large-scale, data-driven characterization of the open YARA rule ecosystem. We have moved beyond anecdotal evidence to quantify its ecosystem dynamics, operational reliability, and threat coverage, revealing a landscape defined by a paradox: it is vast in quantity but operationally unreliable. Our core contribution is this "map" of the ecosystem, which exposes the critical gaps between its perceived value and its practical utility.
The patterns uncovered—from extreme concentration of contributions and widespread maintenance lag to the chasm between static quality and real-world performance—provide a clear mandate for the community. Future work must shift from mere rule creation to intelligent rule curation and lifecycle management. Our findings provide the foundation for this shift, enabling the development of data-driven tools. These tools could automatically assess rule trustworthiness based on ecosystem signals like provenance and update cadence, help practitioners prioritize maintenance by tracking technical lag against evolving threats, and guide rule enrichment by building recommendation systems that identify and fill critical coverage gaps, such as those for underrepresented loaders and stealers.
By moving beyond ad-hoc adoption, this work helps evolve the YARA ecosystem from a loose collection of signatures into a reliable, engineered pillar of modern cyber defense.

\bibliographystyle{ACM-Reference-Format}
  \bibliography{references.bib}

\end{document}